\documentclass[aps,prd,amsmath,groupedaddress,showpacs,showkeys,preprintnumbers,nofootinbib,twocolumn]{revtex4}

 \usepackage[dvips]{graphicx}

\allowdisplaybreaks[4]     

 \renewcommand{\bf}{\bfseries}
 \renewcommand{\it}{\itshape}

 \newcounter{teq}
 \newcommand{\eqlabel}[1]{
   \label{#1}
 }
 \newcommand{\seclabel}[1]{
   \label{#1}
 }

 \newcommand{\td}[2]{\frac{{\rm d} {#1}}{{\rm d} {#2}}}
 \newcommand{\tdil}[2]{{\rm d} {#1} / {\rm d} {#2}}
 \newcommand{\p}{\partial}
 \newcommand{\pd}[2]{\frac{\partial {#1}}{\partial {#2}}}
 \newcommand{\pdil}[2]{\partial {#1} / \partial {#2}}
 \newcommand{\nn}{\nonumber}
 \newcommand{\er}[1]{(\ref{#1})}          

\begin{document}

 \title{Solving the Observer Metric}

 \author{Charles Hellaby}
 \thanks{Charles.Hellaby@uct.ac.za}
 \affiliation{Department of Mathematics and Applied Mathematics, \\
 University of Cape Town, Rondebosch 7701, South Africa}

 \author{ Alnadhief H. A. Alfedeel}
 \thanks{ALFALN001@uct.ac.za}
 \affiliation{Department of Mathematics and Applied Mathematics, \\
 University of Cape Town, Rondebosch 7701, South Africa}

 \date{\today}

 \begin{abstract}

  The analysis of modern cosmological data is becoming an increasingly important task as the amount of data multiplies.  An important goal is to extract geometric information, i.e. the metric of the cosmos, from observational data.  The observer metric is adapted to the reality of observations: information received along the past null cone, and matter flowing along timelike lines.  It provides a potentially very good candidate for developing a general numerical data reduction program.  As a basis for this, we elucidate the spherically symmetric solution, for which there is to date single presentation that is complete and correct.  With future numerical implementation in mind, we give a clear presentation of the mathematical solution in terms of 4 arbitrary functions, the solution algorithm given observational data on the past null cone, and we argue that the evolution from one null cone to the next necessarily involves integrating down each null cone.

 \begin{center}

 {\it Phys. Rev. D, Submitted 11 November 2008, accepted 18 December 2008} \\[1mm]

 \end{center}

 \end{abstract}



 \maketitle

  \section{Introduction}

   Cosmology is all about understanding the observed universe, and since Einstein's field equations are central to that endevour, the primary problem is to determine the cosmic geometry, i.e. the metric, from observations of its matter content.  A full understanding of the dynamics of the universe cannot be separated from understanding its geometry.  Historically, it was very difficult to determine cosmological data with any precision, and the assumption of a homogenous universe, which allowed a simple metric form, was entirely sufficient and indeed very fruitful.  Consequently the problem was reduced to one of finding the best-fit parameter set, rather than determining the metric.  Recent decades have seen a considerable improvement in the quality and quantity of cosmological data, mapping much more accurately the matter distribution and the structures that exist.  Therefore, relaxing the assumption of homogeneity has become an important task.

   The idea of determining the spacetime metric from observational data was first investigated by Kristian and Sachs \cite{KriSac66}, and followed up by Ellis, Stoeger and others in an important series of papers that constitute the ``observational cosmology" (OC) programme \cite{ENMSW85,SNME92,SEN92,SNE92b,SNE92c,MaaMat94,MHMS96,AraSto99,AABFS01,ARS01,RibSto03,AruSto07,AruSto08}.  In this programme, they introduced observer coordinates based on the past null cones of a single observer's worldline, an idea originally due to Temple \cite{Tem38}, because traditional time and space coordinates are not well adapted to cosmological observations.  They also introduced the `fluid-ray tetrad' \cite{SNME92}, including a set of spin coefficients, and their basic equations are derived from this formalism.  Although a general form has been given for the observer metric, work has concentrated on the spherically symmetric case, and to a lesser extent its perturbations.  

   There has also been parallel work relating the Lema\^{\i}tre-Tolman (LT) metric to observations.  One approach uses low-$z$ series expansions \cite{Cel00,TanNam07}, and a more general approach has shown how observational data fully determine an LT model \cite{MHE97,MBHE98,Hel01,Hel06,Lu06,LuHel07,McCHel08}.  
A slightly different formulation can be found in \cite{Ish04}, and an approach based on the characteristic initial value formulation of numerical relativity in \cite{BisHai96}.
All approaches must ultimately be implemented as numerical procedures.

   In this paper we provide a complementary approach to the observational metric.  In the observational cosmology papers, the approach to solving the problem has focussed on using the observational data to analytically determine the metric functions.  Here, with an eventual numerical scheme in mind, our approach emphasises firstly the full formal solution of the field equations for the observer metric, particularly noting the 4 arbitrary functions that emerge in the process and how the evolution is determined by them, and secondly the algorithm for determining the metric from observational data, especially showing how the arbitrary functions are fixed by the data.  The first provides a better understanding of the geometry and dynamics of the model, and the second shows the relationship between the data and the particular characteristics of the solution metric.  We also give the explicit transformation between the OC and LT forms of this metric.

   Although spherical symmetry about the observer is a strong assumption, we regard it as a first step --- a useful and important one --- towards the more general case.  A proper understanding of this simpler case is essential for working with the more general forms of the observer metric.  See \cite{LuHel07,McCHel08} for a more detailed justification.

 \section{Spherically Symmetric Observer Metric}
 \seclabel{SSOM}

We choose coordinates $x^i = (w, y, \theta, \phi)$, and we assume (a) spherical symmetry about the origin, (b) the observer is at the origin, and (c) the $(\theta, \phi)$ surfaces are orthogonal to the $(w, y)$ surfaces.  We work in geometric units.  With these, the metric is
 \begin{align}
   ds^2 & = - A^2 \, dw^2 + 2 A B \, dw \, dy + C^2 \, d\Omega^2 ~,
      \eqlabel{ObsMetric} \\
   \mbox{where}~~~~~~ d\Omega^2 & = d\theta^2 + sin^2\theta \, d\phi^2 ~, \\
   A  = A(w,y) ~& ,~~~~ B = B(w,y) ~,~~~~ C = C(w,y) ~.
 \end{align}
 Here $C$ is an areal radius, and the lack of a $dy^2$ term ensures that the constant $w$ surfaces are null,
 \begin{align}
   dw & = 0 = d\Omega ~~~~\to~~~~ ds^2 = 0 ~.
 \end{align}
 We further assume that the matter is a zero-pressure perfect fluid, comoving with the $y$ coordinate,
 \begin{align}
   T^{ab} & = \rho u^a u^b ~,   \eqlabel{Tab}
 \end{align}
 where the constant $y$, $\theta$, $\phi$ curves have timelike tangent vectors such that
 \begin{align}
   u^a & = \frac{1}{A} \, \delta^a_w ~,~~~~~~
      u_a = (-A, B, 0, 0) ~,~~~~~~
      u^a u_a = -1 ~.
   \eqlabel{ua}
 \end{align}
 Here $\rho$ is the proper density relative to observers on the comoving worldlines, $u^a$.

 We note that the null tangent vector $k_a$ within the constant $w$ surfaces, 
 \begin{align}
   k_a & = \delta_a^w ~,~~~~~~
      k^a = \left( 0, \frac{1}{A B}, 0, 0 \right) ~,~~~~~~
      k^a k_a = 0 ~,
   \eqlabel{ka}
 \end{align}
 must be geodesic since it represents radial light rays, $k^a \nabla_a k^b = 0$, and in fact, for any $K(y)$, $k_a = \delta_a^w K(y)$ is geodesic.  Similarly, the dust particles should follow geodesics, and $u^a \nabla_a u^b = 0$ leads to the two equations
 \begin{align}
   u^w \p_w u^w & = - \left( \frac{A_w}{A} + \frac{B_w}{B} + \frac{A_y}{B} \right)
      (u^w)^2 \\
   0 & = \frac{A}{B} \left( \frac{B_w}{B} + \frac{A_y}{B} \right)
      (u^w)^2 ~,   \eqlabel{yGeodEq}
 \end{align}
 where $A_w = \pdil{A}{w}$, $A_y = \pdil{A}{y}$, etc.  The second of these imposes a restriction of the metric functions,
 \begin{align}
   B_w = - A_y ~,   \eqlabel{Geody}
 \end{align}
 which reduces the first to $(\p_w u^w) / u^w = - (\p_w A) / A$, in agreement with the 
normalisation condition \er{ua}.

   Near the central worldline there is a spherical origin, where $C \to 0$.  The origin conditions for this metric, giving the limiting behaviours of $A$, $B$ and $C$ near an origin, have been presented in several of the observational cosmology papers.
\subsection{Solving the EFEs}
 \seclabel{SolveEFEs}

The Einstein field equations (EFEs) $G^{ab} = \kappa T^{ab} - \Lambda g^{ab}$ for this metric are
 \begin{align}
   G^{ww} & = \frac{2}{A^2 B^2} \left( \frac{A_y C_y}{A C} + \frac{B_y C_y}{B C} - \frac{C_{yy}}{C} \right)
      = \frac{\kappa \rho}{A^2}
      \eqlabel{Gww} \\
   G^{wy} & = \frac{2}{A^2 B^2} \left( \frac{C_{wy}}{C} + \frac{C_w C_y}{C^2} + \frac{A C_y^2}{2 B C^2}
      + \frac{A_y C_y}{B C} - \frac{A B}{2 C^2} \right)\nn \\
& = - \frac{\Lambda}{A B}
      \eqlabel{Gwy} \\
   G^{yy} & = \frac{2}{A^2 B^2} \Bigg( \frac{A_w C_w}{A C} + \frac{B_w C_w}{B C} - \frac{C_{ww}}{C}
         + \frac{A_y C_w}{B C} \nn \\
      &~~~~~~~~~~~~~~~+ \frac{A B_w C_y}{B^2 C} + \frac{A C_w C_y}{B C^2}+ \frac{A A_y C_y}{B^2 C} \nn \\
      &~~~~~~~~~~~~~~~  + \frac{A^2 C_y^2}{2 B^2 C^2} - \frac{A^2}{2 C^2} \Bigg)
         = - \frac{\Lambda}{B^2}
      \eqlabel{Gyy} \\
   G^{\theta\theta} & = \frac{1}{A B C^2} \Bigg( \frac{2 C_{wy}}{C} + \frac{A_{wy}}{A} + \frac{B_{wy}}{B}
         - \frac{A_w A_y}{A^2} \nn \\  
      &~~~~~~~~~~~~~~~ - \frac{B_w B_y}{B^2} + \frac{A C_{yy}}{B C} + \frac{A_{yy}}{B} - \frac{A_y B_y}{B^2} \nn \\
      &~~~~~~~~~~~~~~~ + \frac{A_y C_y}{B C}
         - \frac{A B_y C_y}{B^2 C} \Bigg) = - \frac{\Lambda}{C^2} ~,
      \eqlabel{Gthth}
 \end{align}
 where $\kappa = 8 \pi$, and the conservation equations, $\nabla_b T^{ab} = 0$ are
 \begin{align}
   \nabla_b T^{wb} & = \frac{\rho_w}{A^2}
      + \frac{\rho}{A^2} \left( \frac{2C_w}{C} + \frac{2 B_w}{B} + \frac{A_y}{B} \right) = 0
      \eqlabel{DivTw} \\
   \nabla_b T^{yb} & = \frac{\rho (B_w + A_y)}{A B^2} = 0 ~.
      \eqlabel{DivTy}
 \end{align}

 Not surprisingly, we can obtain \er{Geody} directly from \er{DivTy} since we don't expect the density to be zero.  

 From \er{Gwy} \& \er{Gyy} above we obtain
 \begin{align}
   \frac{A^2 B C}{2} G^{wy} - \frac{A B^2 C}{2} G^{yy} & = \nn \\
     \frac{C_{wy}}{B} - \frac{B_w C_y}{B^2} + \frac{C_{ww}}{A} - \frac{A_w C_w}{A^2} & = 0 ~,
   \eqlabel{Ww}
 \end{align}
 where two terms cancelled because of \er{Geody}.  This can be written as
 \begin{align}
   \pd{}{w} \left( \frac{C_w}{A} + \frac{C_y}{B} \right) = 0 ~,
 \end{align}
 which solves to give
 \begin{align}
   \frac{C_w}{A} + \frac{C_y}{B} = W(y) ~,
   \eqlabel{Wsoln}
 \end{align}
 where $W(y)$ is an undetermined function of integration.  

 Next, from \er{Wsoln} we have
 \begin{align}
   C_y & = B \left( W - \frac{C_w}{A} \right) ~,\nn \\
   C_{wy} & = B_w \left( W - \frac{C_w}{A} \right) - \frac{B}{A^2} \left( A C_{ww} - A_w C_w \right) ~,
 \end{align}
 which combine with the $G^{yy}$ equation \er{Gyy} to give
 \begin{align}
   - B^2 C^2 C_w \, G^{yy} &= 
   \frac{C_w^3}{A^2} + \frac{2 C C_w C_{ww}}{A^2} - \frac{2 C C_w^2 A_w}{A^3} \nn \\ 
&- C_w (W^2 - 1) = C^2 C_w \Lambda ~,
 \end{align}
 where \er{Geody} was used again.  The solution here is
 \begin{align}
   \pd{}{w} \left( \frac{C C_w^2}{A^2} - C (W^2 - 1) - \frac{C^3 \Lambda}{3} \right) & = 0 \\
   \frac{C C_w^2}{A^2} - C (W^2 - 1) - \frac{C^3 \Lambda}{3} & = 2 M(y) ~,
   \eqlabel{Msoln}
 \end{align}
 where $M(y)$ is a second undetermined function of integration. 

 Equations \er{Gww} and \er{Gyy} give the same as the $y$ derivative of \er{Msoln}:
 \begin{align}
   \kappa \rho B W & = (G^{ww} + \Lambda g^{ww}) A^2 B W
      - (G^{wy} + \Lambda g^{wy}) A B C_y \nn \\
   & = - \frac{2 C_w C_y^2}{A B C^2} - \frac{C_y^3}{B^2 C^2} + \frac{C_y}{C^2}
         - C_y \Lambda + \frac{2 B_y C_w C_y}{A B^2 C} \nn \\
      &~~~~~ - \frac{2 C_w C_{yy}}{A B C}+ \frac{2 A_y C_w C_y}{A^2 B C} + \frac{2 B_y C_y^2}{B^3 C}\nn \\
      &~~~~~ - \frac{2 C_y C_{wy}}{A B C} - \frac{2 C_y C_{yy}}{B^2 C}\nn \\
   & = \frac{2 M_y}{C^2} ~,
 \end{align}
 where \er{Wsoln} and its $y$ derivative were used.  Therefore the density is given by
 \begin{align}
   \kappa \rho & = \frac{2 M_y}{C^2 B W} ~,
   \eqlabel{rhoOC}
 \end{align}
 which clearly satisfies \er{DivTw}.  Similarly, the Kretschmann scalar is
 \begin{align}
   K & = R^{abcd} R_{abcd} = \frac{48 M^2}{C^6} + \frac{8 \Lambda^2}{3} - \frac{32 M M_y}{C^5 B W}\nn \\
   & + \frac{12 M_y^2}{C^4 B^2 W^2} + \frac{8 \Lambda M_y}{3 C^2 B W} ~.
   \eqlabel{KretschOC}
 \end{align}

 The solution \er{Msoln} can be re-written as an evolution equation for $C$:
 \begin{align}
   \frac{C_w}{A} & = \pm \sqrt{\frac{2 M}{C} + f + \frac{\Lambda C^2}{3}}\; ~,
      \eqlabel{CwA} \\
   \mbox{where}~~~~~~ f(y) & = W^2 - 1 ~~~~\leftrightarrow~~~~ W = \sqrt{1 + f}\; ~,\eqlabel{f}
 \end{align}
 and the sign depends on whether $C$ is increasing or decreasing with time.  In addition, 
equations \er{Wsoln} and \er{CwA} give	
 \begin{align}
   \frac{C_y}{B} = \sqrt{1 + f}\;
      \mp \sqrt{\frac{2 M}{C} + f + \frac{\Lambda C^2}{3}}\; ~.
   \eqlabel{CyB}
 \end{align}
 Eq \er{CwA} is clearly allied to the LT evolution equation \er{LTev}, except that it contains two unknown functions, $C$ and $A$, so it cannot be solved as is. In this paper, where a function is  transformed between coordinates, we write e.g. $C(w,y)=C(t,r)$, meaning the two forms have the same numerical value at any given event, but they don't have the same functional dependence on their arguments. 
 To solve \er{CwA}, we define $t$ along the worldlines of constant $y$ by
 \begin{align}
   t = \int_{\text{const}\,y} A \, dw ~~~~~~\to~~~~~~
   \pd{}{w} = A \pd{}{t} ~,
   \eqlabel{tDef}
 \end{align}
 which converts \er{CwA} to 
 \begin{align}
   \int \frac{dC}{\pm \sqrt{\frac{2 M}{C} + f + \frac{\Lambda C^2}{3}}\;} = \int dt = t - a(y) ~.
   \eqlabel{C-t-int}
 \end{align}
 In principle this gives us $t(C, y)$ or $C(t, y)$, and introduces $a(y)$, the initial $t$ value at each $y$, as a third free function of integration.  Clearly $t$ is the proper time along each worldline, and the solutions to \er{C-t-int} are identically those of the LT metric.  When $\Lambda = 0$, the solutions for each of the cases $f > 0$, $f = 0$ and $f < 0$ are well known, and are often given parametrically, $\{C(\eta,y), t(\eta, y)\}$.  However, we don't yet have a transformation between $t$ and $w$, since by \er{tDef}
 \begin{align}
   t = \int_{\text{const}\,y} A \, dw ~~~~~~\leftrightarrow~~~~~~ A = t_w ~,
   \eqlabel{Atw}
 \end{align}
 we must know $A$ to calculate $w$ and vice versa.  Now from \er{Geody} we find
 \begin{align}
   B_w = - A_y = - t_{wy} ~~~~\to~~~~ B = - t_y + \beta(y) ~.
   \eqlabel{Bty}
 \end{align}
 The function $\beta(y)$ reflects a freedom in the definition \er{tDef} of $t$,
 \begin{align}
   t ~~\to~~ t + \alpha(y) ~,~~~~~~ \beta ~~\to~~ \beta - \alpha_y ~,
   \eqlabel{tmap}
 \end{align}
 which we shall remove.  We next define
 \begin{align}
   r = r(w,y) = y ~, ~~~~\to~~~~ r_w = 0 ~,~ r_y = 1 ~,   \eqlabel{rDef}
 \end{align}
 and in the next few equations we use $r$ when it is paired with $t$, as in $C_r = \p_r C(t,r)$, but $y$ when paired with $w$, as in $C_y = \p_y C(w,y)$.  By requiring that our $t$ coordinate be orthogonal%
 \footnote{
 Without this, $\beta \neq 0$ and several subsequent equations contain large extra terms.
 }
 to $r$, viz:
 \begin{align}
   0 & = g^{ab} \, (\p_a t) \, (\p_b r)
   = \frac{t_w \, r_y + t_y \, r_w}{A B} + \frac{t_y \, r_y}{B^2}\nn \\
   & = \frac{A + 0}{A B} + \frac{(\beta - B)}{B^2}
   ~~~~\to~~~~ \beta = 0 ~,
   \eqlabel{beta0}
 \end{align}
 we reduce the freedom in $t$ to a constant translation, i.e. $\alpha$ is a constant, but henceforth we shall drop it from our equations.  Evidently, then, $B$ is the negative of the rate of variation of proper time with respect to $y$ down the past null cone.  Even though we know $t(C,y)$, we can't calculate $t_y$ unless we know how to hold $w$ constant.  Now the transformation between $C(w,y)$ and $C(t,r)$ allows us to write
 \begin{align}
   C_y & = C_t t_y + C_r r_y = - C_t B + C_r ~,
      \eqlabel{CyTransf}
 \end{align}
 where $C_t \equiv \pdil{C}{t}$ and $C_r \equiv \pdil{C}{r}$.  Combining
  \er{CyTransf} with \er{CyB} leads to 
 \begin{align}
   B & = \frac{C_r}{\sqrt{1 + f}\;} ~,
   \eqlabel{B}
 \end{align}
 and since $C(t,r)$ is known, this gives us $B(t,r)$.  Using \er{Bty}, \er{beta0} and \er{B} we obtain the differential equation
 \begin{align}
   t_y & = \frac{- C_r}{\sqrt{1 + f}\;} ~.
      \eqlabel{tyDE}
 \end{align}
 This equation specifies how much $t$ changes for a given $y$ change, when $w$ is constant, so it may be integrated down the null cones, i.e. along constant $w$, from the origin outwards, giving $t(w, y)$.  The boundary conditions, fixed say at the origin $y = 0$, give us a 4th undetermined function, $\gamma(w) = t(w, 0)$.  Actually, this fixes the variation of $w$ with respect to $t$, as $t$ is fixed by integrating \er{C-t-int}.  An obvious choice is $w = t|_o = \gamma$.  Having solved \er{tyDE}, we can then convert $C(t, r)$ to $C(w, y)$ using
 \begin{align}
   C(t, r) ~~\mbox{plus}~~ t(w,y) ~~~~\to~~~~ C(w,y) = C(t(w,y), y) ~,
   \eqlabel{GettingCofwy}
 \end{align}
 and we finally determine $A(w,y)$ from \er{Atw}, and $B(w,y)$ from \er{Bty}, or possibly \er{B}.  The algorithm for calculating the model evolution is detailed in section \ref{alg}.

 Having completed the solution, we see that this metric has 4 arbitrary functions --- $f(y)$, $M(y)$, $a(y)$ \& $\gamma(w)$ --- of which $\gamma(w)$ represents a freedom to rescale $w$ that is most naturally set to $\gamma = w$.  The physical meanings of $f$, $M$ and $a$ are exactly as in the LT model; they represent two physical relationships plus a freedom to rescale $y$.  

 \section{Observable Quantities}
 \seclabel{OQ}

 The primary observables for cosmological sources are redshift $z$, angular diameter $\delta$, apparent luminosity $\ell$, and number density in redshift space $n$.  Associated with each of $\delta$, $\ell$, and $n$ is a source property, true diameter $D$, absolute luminosity $L$, and mass per source $\mu$, that is needed to relate the observations to the theoretical model.  The observables all depend on redshift $z$, and since sources evolve with time, so do the source properties.  We assume that the source properties are known from a combination of observations and source evolution theories.

 Let $w_0$ label the past null cone (PNC) of present day observations by the central observer at $(t, r) = (t_0, 0) \equiv (w, y) = (w_0, 0)$, and let the evaluation of any quantity $Q(w,y)$ on this PNC be denoted $\hat{Q} = [Q]_\wedge = Q(w_0, y)$.  We assume that emitters  follow comoving worldlines $y_e$, and the observer is at the central worldline, $y_o = 0$.  Let the evaluation of a quantity at the observer and the emitter be denoted $Q_o(w) = Q(w, 0)$ and $Q_e(w) = Q(w, y_e)$ respectively. However we will often drop the subscript $e$.

 The redshift of comoving sources on that null cone is given by the ratio of the light oscillation periods $T$ measured at the observer, $o$, and the emitter, $e$,
 \begin{align}
   (1 + z) & = \frac{T_o}{T_e} = \frac{\hat{A}_o \, dw}{\hat{A}_e \, dw} ~~~~\to~~~~
   \hat{A} = \frac{\hat{A}_o}{(1 + z)} ~,
   \eqlabel{zA}
 \end{align}
 and we can put $\hat{A}_o = 1$ since $\hat{A}_o = A(w_0, 0) = \p_w \gamma|_{w_0} = 1$ is the natural choice.

   The diameter distance of a source is the true diameter $D$ divided by the angular diameter $\delta$, and in a spherical metric it corresponds to the areal radius evaluated on the PNC, i.e.
 \begin{align}
   \frac{D}{\delta} = d_D = \hat{C}_e ~.   \eqlabel{dD}
 \end{align}
 Similarly, if the absolute luminosity of a source is $L$, the apparent luminosity is $\ell$ (or $m$ and $\tilde{m}$ the apparent and absolute magnitude), and $d_{10}$ is 10 parsecs, then the luminosity distance is
 \begin{align}
   d_L = \sqrt{\frac{L}{\ell}}\; \, d_{10} = 10^{(m - \tilde{m})/5} \, d_{10} ~.   \eqlabel{dL}
 \end{align}
 By the reciprocity theorem \cite{Eth33,Penr66,Ell71}, $d_L$ may be converted to $d_D$ using $z$.
 \begin{align}
   d_D = {d_L}{(1 + z)^2} ~.   \eqlabel{rcprct}
 \end{align}

 In redshift space, $(z, \theta, \phi)$, let $n(z)$ be the density of sources, that is the number per steradian per unit redshift interval%
 \footnote{Thus this $n$ is different from the $n$ used in the OC programme, which is number density on a constant time slice, see section \ref{ROCP}.}%
 .  Suppose that there are $dN$ sources in solid angle $d\omega = \sin\theta \, d\theta \, d\phi$ between redshift $z$ and $z + dz$, and that $\mu(z)$ is the mean mass per source%
 \footnote{For a treatment with a variety of source types, see \cite{Hel01}.}%
 , then the mass in that volume element of redshift space is
 \begin{align}
   d{\cal M} = \mu \, dN = \mu \, n \, d\omega \, dz ~.
 \end{align}
 The proper 3-volume enclosing these sources at the time of emission, as measured by comoving observers $u^a$, is spanned by
 \begin{align}
   dx_1^a = \delta^a_y \, dy ~,~~~~ 
   dx_2^a = \delta^a_\theta \, d\theta ~,~~~~ 
   dx_3^a = \delta^a_\phi \, d\phi ~,
 \end{align}
 and evaluates to
 \begin{align}
   d^3v& = \eta_{abcd} \, u^a \, dx^b_1 \, dx^c_2 \, dx^d_3
      = \sqrt{|g|}\; \, \epsilon_{0123} \, u^0 \, dx^1_1 \, dx^2_2 \, dx^3_3 \nn \\
   & = \sqrt{|-A^2 B^2 C^4 \sin^2 \theta|} \, \frac{1}{A}\, dy \, d\theta \, d\phi \nn \\
   & = B C^2 \sin\theta \, d\theta \, d\phi \, dy
      = B C^2 \, d\omega \, dy ~,
 \end{align}
 so that the mass in this fluid element is
 \begin{align}
   d{\cal M} & = \rho \, B \, C^2 \, d\omega \, dy ~.
 \end{align}
 Clearly we have the following relationship between $n$ and $\hat{\rho}$,
 \begin{align}
   \mu \, n = \hat{\rho} \, \hat{B} \, \hat{C}^2 \, \td{y}{z} ~.
   \eqlabel{munrho}
 \end{align}

 The apparent horizon is where $C$ is maximum on any given constant $w$ cone,
 \begin{align}
   C_y = 0 ~.
 \end{align}
Now if the metric (\ref{ObsMetric}) is to be regular, and the density \er{rhoOC} and Kretschmann scalar \er{KretschOC} finite at such a point,  then $B$ must be non-zero. This would need the upper sign in \er{CyB} --- that is, the local matter-shells are expanding, $\dot{C} > 0$ --- and 
 \begin{align}
   \sqrt{\frac{2 M}{C} + f + \frac{\Lambda C^2}{3}}\; & = \sqrt{1 + f}\;\nn \\
    ~~~~\to~~~~ 6 M + \Lambda C^3 - 3 C & = 0 ~,\eqlabel{ObsAHCond}
 \end{align}
 along this locus. When $\Lambda  = 0 $ this simplifies to $C = 2 M$. As shown in \cite{Hel06,McCHel08}, this locus has considerable observational significance.  See also \cite{AruSto07}.

 \section{Determining the Solution from Observational Data}
 \seclabel{DSOD}

 Given the above observational data on the past null cone, that is $\hat{A}(z)$, $\hat{C}(z)$, and $\mu n(z)$, the solution process must determine the arbitrary functions $f$, $M$, $a$ and $\gamma$.  Knowing these enables all spacetime quantities to be calculated and evolved, via the results of section \ref{SolveEFEs}.  We envisage a numerical solution, so the emphasis here is on laying out a solution algorithm, rather than on formal integrals and functional dependence.

 \subsection{Gauge choices}
 \seclabel{GC}

   The observational data must determine the physical properties of the model, but cannot restrict the coordinate freedoms.  Therefore we will have to make some gauge choices in order to effect the solution.  Firstly, we set $w = t$ along the central worldline, which implies
 \begin{align}
   \gamma(w) & = w ~~~~\to~~~~ A(w,0) = t_w(w, 0) = 1 ~~~~\to~~~~\nn \\
   \hat{A}_o &= 1 ~~~~\to~~~~ \hat{A} = \frac{1}{(1 + z)} ~.
 \end{align}
 Secondly, we need to set the freedom in the $y$ coordinate.  We consider two options below.  The `LT' option specifies
 \begin{align}
   \hat{t}_y = -1 ~~~~\to~~~~ \hat{B} = 1 ~,
   \eqlabel{BtyLT}
 \end{align}
 as in many LT approaches.  The OC papers choose $A(w_0, y) = B(w_0, y)$ on the PNC, so in the `OC' option we choose
 \begin{align}
   \hat{B} = \hat{A} = \frac{1}{(1 + z)} ~~~~\to~~~~ \hat{t}_y = \frac{- 1}{(1 + z)} ~.
   \eqlabel{BtyOC}
 \end{align}

 \subsection{DE for $y(z)$}
 \seclabel{DEy}

 The coordinate $y$ is of course not observable, but we have to determine it first.  We define
 \begin{align}
   \varphi = \td{y}{z} ~,
   \eqlabel{phiDef}
 \end{align}
 and along the PNC we re-write our equations in terms of $z$ derivatives rather than $y$ derivatives; for any quantity $Q(w,y)$,
 \begin{align}
    Q_y = \frac{Q_z}{\varphi} ~~~~\mbox{and}~~~~ Q_{yy} = \frac{Q_{zz}}{\varphi^2} - \frac{Q_z \varphi_z}{\varphi^3} ~.
 \end{align}
 Evaluating \er{Gww} on the PNC, and using \er{munrho}, we find
 \begin{align}
   \frac{\hat{A}_z \hat{C}_z}{\hat{A} \hat{C}} + \frac{\hat{B}_z \hat{C}_z}{\hat{B} \hat{C}} 
   - \frac{\hat{C}_{zz}}{\hat{C}} + \frac{\hat{C}_z \varphi_z}{\hat{C} \varphi}
   - \frac{\kappa \mu n \hat{B} \varphi}{2 \hat{C}^2} = 0 ~,
 \end{align}
 which, upon substituting for $\hat{A}_z/\hat{A}$ from \er{zA}, leads to
 \begin{align}
   \varphi_z = \varphi \left( \frac{1}{(1 + z)} - \frac{\hat{B}_z}{\hat{B}} 
   + \frac{\hat{C}_{zz}}{\hat{C}_z}
   + \frac{\kappa \mu n \hat{B} \varphi}{2 \hat{C} \hat{C}_z} \right) ~,
   \eqlabel{phiDEB}
 \end{align}
 where all derivatives are now total derivatives along the PNC.

   At this point we must fix the gauge in order to freeze out the coordinate freedom.  The two options given above each convert \er{phiDEB} to an ODE for $\varphi(y)$ completely in terms of observables:
 \begin{align}
   \mbox{OC:}~~~~~~ (\varphi_1)_z & = \varphi_1 \left( \frac{2}{(1 + z)} 
      + \frac{\hat{C}_{zz}}{\hat{C}_z}
      + \frac{\kappa \mu n \varphi_1}{2 (1 + z) \hat{C} \hat{C}_z} \right)
      \eqlabel{phiDE1} \\
   \mbox{LT:}~~~~~~ (\varphi_2)_z & = \varphi_2 \left( \frac{1}{(1 + z)}
      + \frac{\hat{C}_{zz}}{\hat{C}_z} + \frac{\kappa \mu n \varphi_2}{2 \hat{C} \hat{C}_z} \right) ~.
      \eqlabel{phiDE2}
 \end{align}
 Integrating \er{phiDE1} or \er{phiDE2} followed by \er{phiDef} yields $\varphi_i(z)$ and
 \begin{align}
   y_i(z) = \int_0^z \varphi_i(z) \, dz ~.
      \eqlabel{yzint}
 \end{align}
 This allows us to convert between functions of $z$ and functions of $y$ on the PNC.

 \subsection{DE for $M(z)$ \& $W(z)$}
 \seclabel{DEMW}

 From \er{rhoOC} on the PNC and \er{phiDEB} we obtain
 \begin{align}
   M_z & = \frac{\kappa \mu n W}{2} ~,
   \eqlabel{MzDE}
 \end{align}
 where $W$ is found by putting \er{CyB} on the PNC,
 \begin{align}
   W = \frac{\hat{B} \varphi}{2 \hat{C}_z} \left( 1 - \frac{2 M}{\hat{C}}
   - \frac{\Lambda \hat{C}^2}{3} \right) + \frac{\hat{C}_z}{2 \hat{B} \varphi} ~.
   \eqlabel{WMC}
 \end{align}
 The two gauge choices give
 \begin{align}
   \mbox{OC:}~~~~~~ W & = \frac{\varphi_1}{2 \hat{C}_z (1 + z)} \left( 1 - \frac{2 M}{\hat{C}}
      - \frac{\Lambda \hat{C}^2}{3} \right) \nn \\ 
     & + \frac{\hat{C}_z (1 + z)}{2 \varphi_1}
      \eqlabel{WMC1} \\
   \mbox{LT:}~~~~~~ W & = \frac{\varphi_2}{2 \hat{C}_z} \left( 1 - \frac{2 M}{\hat{C}}
      - \frac{\Lambda \hat{C}^2}{3} \right) + \frac{\hat{C}_z}{2 \varphi_2} ~.
      \eqlabel{WMC2}
 \end{align}
 Together \er{MzDE} and \er{WMC1} or \er{WMC2} constitute an ODE for $M(z)$ that also generates $W(z)$.  Note that \er{WMC1} requires $\varphi_1$ from \er{phiDE1} and \er{WMC2} requires $\varphi_2$ from \er{phiDE2}.  Technically, this is a first order linear inhomogeneous ODE for $M$, so the formal solution is well known.  In practice, the two integrals involved would both have to be done numerically, so it is less work to solve the ODE directly in parallel with \er{phiDEB} and \er{phiDef}, using say a Runge-Kutta method.
 
 \subsection{Obtaining $a(z)$}
 \seclabel{Oa}

 From \er{Bty} and \er{beta0} on the PNC we get
 \begin{align}
   \hat{t}_z = - \varphi \hat{B} ~,
   \eqlabel{tzB}
 \end{align}
 which, in the OC \& LT gauges, simplifies to the ODEs
 \begin{align}
   \mbox{OC:}~~~~~~ \hat{t}_z & = \frac{- \varphi_1}{(1 + z)}
      \eqlabel{tzB1} \\
   \mbox{LT:}~~~~~~ \hat{t}_z & = - \varphi_2 ~,
      \eqlabel{tzB2}
 \end{align}
 thus giving the worldline proper time, $\hat{t}(z)$ or $\hat{t}(y)$, on the PNC.  The appropriate $\varphi_i$ must be used in each equation.  
From \er{C-t-int} on the PNC, we write
 \begin{align}
   \int_0^{\hat{C}} \frac{dC}{\pm \sqrt{\frac{2 M}{C} + f + \frac{\Lambda C^2}{3}}\;} = \tau ~,
   \eqlabel{tau}
 \end{align}
 where $f$ is given in \er{f}  and $\tau$ is the proper time from the bang to the PNC along the matter worldlines, and after performing the integral at each $z$, we calculate
 \begin{align}
   a(z) = \hat{t}(z) - \tau(z) ~.   \eqlabel{athtau}
 \end{align}

 We now have $M$, $W = \sqrt{1 + f}\;$, and $a$, and we've also used up the freedom to rescale $y$ by directly or indirectly fixing $t_y$.  The only undetermined function is $\gamma(w)$, though we have already fixed $\hat{A}_o = t_w(w_o, 0) = 1$, and we should extend this to
 \begin{align}
   \gamma = w ~.
 \end{align}

 \subsection{Evolving off the PNC}
 \seclabel{EoPNC}

   In principle, no amount of data on the PNC is sufficient to determine the future evolution of any part of the spacetime, because new information can arrive along succeeding incoming light rays.  But since we have already assumed a dust equation of state, in order to get the arbitrary functions, it is not unreasonable to expect the worldlines continue their dust evolution into the future.  This same assumption is tacitly made when fitting a Robertson-Walker model to observational data.

   Away from the PNC, $z$ is no longer a useful variable, and we should rather use $y$.  Also, the gauge choices don't give $B$ off the PNC, so gauge-specific equations such as \er{phiDE1} or \er{tzB2} are not applicable.  One may determine the full evolution of $C$ using the algorithm below, based on the solution of section \ref{SolveEFEs}.  In addition, \er{Gww} together with \er{rhoOC} provides a cross-check on the calculated propagation of the metric components.

   The remaining question is whether, given that we have initial data for $\hat{C}$, $\hat{B}$ \& $\hat{A}$ on the PNC, there is a better way to evolve $C$ than integrating over the entire $(t,r)$ domain twice, first calculating $C(t,r)$ and then finding $t(w,y)$ and converting to $C(w, y)$.  Can we integrate directly with respect to $w$, giving $C(w, y)$ straight off?  This would be especially important if there were detectable time evolution in cosmological observables.  The key difficulty is that we don't know any of $A$, $B$ or $C$ away from the PNC and the central worldline, and though we have direct evolution equations for $B_w$ and $C_w$, there isn't one for $A_w$.

 Once the arbitrary functions $W$, $M$ and $a$ are known, the observational data plus the gauge choices give us all of $A$, $B$ \& $C$ on an initial constant $w$ null cone, $w_0$.  For clarity of argument, consider Euler integration.  Evolution equations for $C$ and $B$ follow from \er{CwA} and \er{Geody},
 \begin{align}
   C_{i+1}  & = C_i + (C_w)_i \, dw ~,~~~~~~
      C_w   = A V ~,\nn \\
        & V = \pm \sqrt{\frac{2 M}{C} + W^2 - 1 + \frac{\Lambda C^2}{3}}\; ~,
      \eqlabel{Cw} \\
   B_{i+1} & = B_i + (B_w)_i \, dw ~,~~~~~~
      B_w  = - A_y ~,
      \eqlabel{Bw}
 \end{align}
 but the difficulty is finding an evolution equation for $A$.  For example, \er{Wsoln}, in the form
 \begin{align}
   A_{i+1} & = \frac{(C_w)_{i+1}}{W - (C_y)_{i+1} / B_{i+1}} = \frac{A_{i+1} V_{i+1}}{W - (C_y)_{i+1} / B_{i+1}} ~,
      \eqlabel{Ai+1}
 \end{align}
 does not help because $A_{i+1}$ cancels out.  The $yy$ EFE contains $A_w$, but as soon as we substitute $C_{ww} = A_w V + A V_w$, then $A_w$ vanishes from the equation.  The best we can do is to put $B_w = - A_y$ and $C_w = A V$ and $C_{ww} = A_w V + A V_w$ in the $wy$ EFE, to obtain an expression for $A_y/A$ that is free of $w$ derivatives; but this requires an integration along constant $w$, and is at least as much work as solving \er{tyDE}.  Similarly the $\theta\theta$ EFE gives an expression for $\p_w(A_y/A)$.

 Combining \er{CwA} and \er{CyB} with \er{Geody} eliminates $A$ \& $B$, but leads to a 
second order non-linear PDE for $C$ that depends only on $M$, $f$ (or $W$), and their $y$ derivatives.  This would also require a double numerical integration over the $(w, y)$ space.  We have not been able to cast it in a simpler form, and we don't regard it as a better alternative, so we don't write it out here.  

Evidently there is no direct integration off the PNC along the constant $y$ worldlines, though it should be possible to program the numerical integration as a single sweep across the spacetime.

 \subsection{The Algorithm}
 \seclabel{alg}

   The procedure for obtaining the observer metric from observational data may be presented as a two-part algorithm, the first for obtaining the undetermined functions from the data, and the second for calculating the model evolution from the functions.  The arbitrary functions $M$, $W = \sqrt{1 + f}\;$ and $a$ are obtained as follows:
 \begin{itemize}

 \item   Assume the following observational data for a large number of sources on the PNC: \\
 ${}$ \hspace*{5mm}\parbox[t]{14cm}{
 redshift $z$, \\
 apparent luminosity $\ell$ and absolute luminosity $L$, \\
 (and/or angular diameter $\delta$ and true diameter $D$), \\
 number density of sources in redshift space $n$\\
 and mass per source $\mu$. } \\
 From these calculate diameter distance $\hat{C} = d_D(z)$ using \er{dD} and \er{dL}, and redshift space mass density $\mu n(z)$.

 \item   Make a gauge choice, as in \S \ref{GC}, which fixes $\hat{B}$ and $\hat{t}_y$.

 \item   Integrate down the PNC one of equations \er{phiDEB}/\er{phiDE1}/\er{phiDE2}, as appropriate to the gauge choice, which gives $\varphi(z)$, then integrate $\varphi(z)$ as in \er{yzint} to produce $y(z)$ $\to$ $z(y)$.

 \item   Integrate \er{MzDE} with the appropriate choice of \er{WMC}/\er{WMC1}/\er{WMC2} down the PNC to calculate $M(z)$ and $W(z)$ $\to$ $M(y)$ \& $W(y)$.

 \item   Integrate the relevant choice of \er{tzB}/\er{tzB1}/\er{tzB2} to give the time on the PNC $\hat{t}(z)$; integrate \er{tau} along each constant $y$ worldline, producing $\tau(z)$ the proper time from bang to null cone; then calculate the bang time $a(z)$ from \er{athtau} $\to$ $\hat{t}(y)$ \& $a(y)$.

 \end{itemize}
   Having found the 3 arbitrary functions, the evolution of the model is determined as follows:
 \begin{itemize}

 \item   Use equation \er{C-t-int} and integrate up and down each matter worldline to evaluate $t(C,r)$ $\to$ $C(t,r)$ everywhere ($r = y$).  Initial conditions are provided on the PNC by $\hat{C}(z(y))$ and $\hat{t}(y)$.  In practice, this could be done in the same step as the $\tau$ integration above.

 \item   Choose the gauge function $\gamma(w) = t(w, 0)$ to fix $w$ all along the central worldline.

 \item   Knowing $C(t,r)$, calculate $C_r$ everywhere; and hence find $B(t,r)$ everywhere from \er{B}.

 \item   From each $w$ on the central worldline, integrate equation \er{tyDE} to trace the $(t, y)$ locus of its past null cone, allocating each point the same $w$, thus obtaining $w(t, y)$ $\to$ $t(w, y)$.  

 \item   Calculate $C(w, t) = C(t(w, y), y)$ and $B(w, t) = B(t(w, y), y)$ everywhere, as shown in \er{GettingCofwy}.

 \item   Differentiate $t(w,y)$ to find $A(w,y)$ according to \er{Atw}.

 \end{itemize}
 The above steps are written for clarity rather than numerical efficiency.  In coding it, certain steps may be combined.  As explained in \cite{LuHel07,McCHel08}, the neighbourhoods of the origin, the bang, parabolic worldlines, and the maximum in $d_D$ require special numerical treatment.  
 
 \section{Relationship to Other Work}
 \seclabel{ROCP}

   This solution must obviously be a version of the LT metric, so it would be useful to see the transformation.  We give this in appendix \ref{TLTNCC}.

We here compare the present paper with earlier work, particularly noting any differences, and we give the conversion between the different notations that have been used

 The notation for coordinates, $(w, y, \theta, \phi)$; and metric functions $A$, $B$, $C$ is common to all the papers, as are those for the redshift $z$ and the primary PNC $w_0$.  In the OC papers, the $w$ and $y$ partial derivatives are written $\dot{C} = \p_w C = C_w$, $C' = \p_y C = C_y$, etc.

   Concerning \cite{SEN92}, we note there is a factor of $\kappa$ missing on the right of their equations (13), or it has been absorbed into the definitions of $\mu$, $\mu_0$ and $p$.  We will assume the latter.  They use a completeness factor $F$ --- the fraction of sources that are actually counted.  Whereas survey completeness is a significant concern, in our paper we have assumed it has been corrected for.  (The effect of systematic errors was investigated in \cite{McCHel08}.)  Combining our \er{DivTw} and \er{DivTy} gives
 \begin{align}
   \frac{\rho_w}{\rho} & = - \left( \frac{2C_w}{C} + \frac{B_w}{B} \right) 
   ~~~~~~\to~~~~~~ \nn \\
\rho & = \frac{\overline{\rho}(y)}{C^2 B}
   ~,~~~~~~ \overline{\rho} = \frac{2 M_y}{\kappa W}
 \end{align}
 which is (21a) in \cite{SEN92}.  As noted in \cite{MHMS96}, eq (30) of  \cite{SEN92} does not hold, so eqs their (34)-(45) are incorrect.

   In \cite{MHMS96}, they effectively obtain orthogonality of the constant $t$ and $y$ surfaces in their (13) by comparing the matter flow lines in the LT and OC metrics.  Their variable $N_*(y)$ is never interpreted --- it is $8 \pi$ times the total gravitational mass $M$ within shell $y$, divided by the mean galaxy mass.  Since $M$ contains both the integrated rest mass and the curvature, $N_*$ is not proportional to the number of galaxies (except near the centre).  Item 4 in the corrigenda is misleading; to get that solution, put $u = F - v$ into $v^2 + 2 u v = 1 - mN_*/C$, multiply through by $C$, rearrange, differentiate with respect to $y$, and use the derivative of (27), giving
 \begin{align}
   (2 C v F)' + m F N' = (C v^2 + C)' ~,
 \end{align}
 which can be evaluated on the PNC and integrated to give $F$.  Note the prime in their paper (and in this paragraph) is $\pdil{}{y}$ while $w$ is held constant.  This must be remembered when taking the prime derivatives on the right of their (29), where functions are expressed in terms of $t$, $y$, and parameter $\Gamma$.  In fact, we recommend integrating the negative of the left hand side of (29), to get the null cone path $t(y)$, and then determining $T$ rather than $T'$.  Thus the derivative of the parametric solution, which includes calculating $\Gamma'$, is not needed.  There seem to be some oddly placed factors of $4 \pi$: if their (17) and the equation below their (10) are correct, then their $\rho$ is $4 \pi$ times the density.  Also their (17) and (27) imply their $mN_*$ is our $M$, whereas their (25) implies it is our $2M$.  We will assume the latter, as their equations involving $\rho$ do not play a significant role.  Otherwise that paper (with corrigenda) is basically correct, though a little circuitous, and the propagation of the solution off the observer's PNC in observer coordinates is not discussed.  

   In \cite{ARS01}, $\mu$ and $\mu_0$ also contain absorbed factors of $\kappa$.  The function of integration $l(y)$ in their (47) corresponds to $\beta(y)$ in our \er{Bty}, but, without the orthogonality condition, they haven't set it to zero.  In steps 2 and 3 of their integration procedure, they argue that $A(w, y)$ must have the same functional form as $A(w_0, y)$, though they acknowledge that this does not mean simple replacement of $w_0$ by $w$.  In their FLRW example, they use the central conditions plus homogeneity to obtain $1 \, \to \, w/w_0$ in the formula for $A$.  Unfortunately, we cannot see how this can be implemented in general, away from the origin, when we aren't assuming homogeneity.  

 The table below summarises the correspondence between the different notations.

 \[
 \setlength{\tabcolsep}{2mm}
 \begin{tabular}{|l|l|l|l|l|l|}
 \hline
 Here & In \cite{SEN92} & In \cite{MHMS96} & In \cite{ARS01} \\
 \hline
 \hline
   \rule{0mm}{4.5mm}%
   $\hat{C}$
   & $r_0$
   & $d_A$
   & $r_0$ \\
 \hline
   $\mu$
   & $M$
   & $4 \pi m$
   & $m$ \\
 \hline
   \rule[-2mm]{0mm}{6mm}%
   $\rho$
   & $\frac{\mu}{\kappa}$
   & $\frac{\rho}{4 \pi}$
   & $\frac{\mu}{\kappa}$ \\
 \hline
   \rule[-2.5mm]{0mm}{6mm}%
   $\frac{\rho}{\mu}$
   & $n$
   & $n$
   & $n$ \\
 \hline
   \rule[-2.5mm]{0mm}{6mm}%
   $\frac{\kappa \mu n}{\hat{C}^2}$
   & $M_0 F$
   &
   & $M_0 J$ \\
 \hline
   \rule[-2mm]{0mm}{6.5mm}%
   $n$
   & $\frac{F M_0 r_0^2}{\kappa M}$
   & $\frac{N'(\tdil{y}{z})}{4 \pi}$
   & $\frac{J M_0 r_0^2}{\kappa m}$ \\
 \hline
   \rule[-2mm]{0mm}{6.8mm}%
   $\kappa \overline{\rho} = \frac{2M_y}{W}$
   & $\mu_0$
   &
   & $\mu_0$ \\
 \hline
   \rule[-2mm]{0mm}{6mm}%
   $M$
   & $-\omega_0$
   & $\frac{m N_*}{2}$
   & $-\omega_0$ \\
 \hline
   \rule[-2mm]{0mm}{6mm}%
   $-\frac{M}{C^3}$
   & $\omega$
   &
   & $\omega$ \\
 \hline
   \rule[-2.5mm]{0mm}{6mm}%
   $\frac{8 \pi M}{\mu}$
   & 
   & $N_*$
   & \\
 \hline
   $W$
   & $W$
   & $F$
   & $W$ \\
 \hline
   $f$
   & 
   & $-k f^2$
   & \\
 \hline
   $a$
   & 
   & $-T$
   & \\
 \hline
   $\beta$
   & 
   &
   & $l$ \\
 \hline
   $\gamma$
   & $A(w, 0)$
   &
   & $A(w, 0)$ \\
 \hline
 \end{tabular}
 \]
 \section{Conclusions}
 \seclabel{Concl}

 We have presented two methods of solving the spherical observer metric; firstly a formal solution of the EFEs in terms of arbitrary functions, and secondly a procedure for determining the arbitrary functions and the metric evolution from observational data.  Both are given in a clear, step by step manner.  We have been precise about where the gauge choices are made, and have considered two possible sets.  Our main aim was to be able to lay out a solution algorithm that could be coded for numerical implementation.  This naturally divides into to two distinct stages: determining the arbitrary functions from given observational data, and determining the spacetime evolution from known arbitrary functions.  These are given in \S \ref{alg}.  The understanding thus gained will be useful for generalisations away from spherical symmetry.

   We emphasise that, if we are given observational data, then they fix the arbitrary functions, and we are only free to make the gauge choices that pin down the coordinate freedoms.  On the other hand, if we choose all the arbitrary functions, then the observational relations are already fixed, and there's no room to fit observations.  We find previous solution methods have not so clearly distinguished the two, and can be hard to follow.  We also find those methods were sometimes ambiguous about the functional dependence of their solution functions, especially when extending the solution off the PNC, so that it was not always apparent how to proceed with calculations.

   The OC papers all start from the fluid ray tetrad equations, though a number of different solution methods and notations are used.  Our main contribution is in clarifying how the model evolution is to be calculated.  Although the error in \cite{SEN92} was corrected in \cite{MHMS96}, the latter does not really address model evolution, and the method suggested in \cite{AraSto99,ARS01} may not be practicable.  Also, unlike \cite{MHMS96,AraSto99} our method does not need to refer to a known equivalent metric to restrict the solution

   In the formal solution, our approach has a lot in common with previous approaches up to \er{Bty}, but thereafter, up to the final solution \er{GettingCofwy}, it is new.  In the solution from observations, our method is similar to others in sections \ref{DEy} and to some extent \ref{Oa} where $y(z)$ and $a(z)$ are found, but differs in sections \ref{DEMW} and \ref{EoPNC} where $M(z)$ and $W(z)$ are found and the evolution off the PNC is discussed.  In particular, the demonstration that the evolution from one constant $w$ null cone to the next necessarily involves an integration down the null cone at each step, is new.  The orthogonality condition leading to $\beta = 0$ is important for the solution from scratch, but up to now has only been obtained indirectly by comparing with the LT metric.  

 The available data on galaxy observations only extends to a finite redshift $z$.  This is well suited to the algorithms presented here, which involve integrations along the null cones outwards from the centre, and integrations along the constant $y$ worldlines.  The LT functions are fully determined within the range of reliable data.  Even if quite large scale inhomogeneity is discovered, it is still possible the cosmos approaches homogeneity on even larger scales.

In \cite{MHE97} it was shown that any reasonable $\hat{C}(z)$ and $\mu n(z)$ derived from observations could be fitted by an LT model with zero $\Lambda$.  While this is true for nearly all $z$, \cite{Hel06} pointed out that the maximum in the diameter distance is an exception.  The properties of this particular locus allow the mass $M$ to be calculated independently of \er{MzDE}.  With perfect observations, this cross-check would allow a determination of $\Lambda$.  In practice, with real observations, there would be systematic errors, and as shown in \cite{LuHel07,McCHel08} this cross-check may instead be used to detect and correct for systematic errors.  Thus, one would need more than the observations considered here to distinguish non-zero $\Lambda$ from inhomogeneity.

 \setcounter{secnumdepth}{0}
 \section{Acknowledgements}
 \setcounter{secnumdepth}{2}

 CH thanks South Africa's National Research Foundation (NRF) for a grant.  
 AHAA thanks the NRF and the African Institute for Mathematical Sciences (AIMS) for bursaries.
 We thank Bill Stoeger for many very useful discussions, and George Ellis for helpful comments.


 \appendix

 \section{The Transformation of LT to Null-Comoving Coordinates}
 \seclabel{TLTNCC}

 The LT (Lema\^{\i}tre-Tolman) metric is \cite{Lem33,Tol34}
 \begin{align}
   ds^2 & = - dt^2 + \frac{(R')^2}{1 + f} \, dr^2 + R^2 \, d\Omega^2 ~,
 \end{align}
 where $R = R(t, r)$ and $R' = \pdil{R}{r}$.  It depends on 3 arbitrary functions, $f = f(r)$, $M = M(r)$ and $a = a(r)$.  The matter is comoving and has zero pressure.  The evolution equation is
 \begin{align}
   \dot{R}^2 & = \frac{2 M}{R} + f + \frac{\Lambda R^2}{3} ~,
   \eqlabel{LTev}
 \end{align}
 where $\dot{R} = \pdil{R}{r}$, and the density is given by
 \begin{align}
   \kappa \rho & = \frac{2 M'}{R^2 R'} ~.
   \eqlabel{LTrho}
 \end{align}
 The arbitrary functions each have a physical meaning; $f$ gives the deviation of the constant $t$ 3-spaces from flatness, and also gives twice the energy per unit mass of the dust particles; $M$ gives the gravitational mass within the comoving shells of constant $r$, and $a$ gives the local time of the big bang on each constant $r$ worldline.

 We propose the transformation
 \begin{align}
   t & = t(w, y) ~,~~~~~~ r = y\nn \\
   ~~~~\to~~~~ J & = \pd{(t,r)}{(w,y)}  
      = 
      \begin{pmatrix} 
         t_w & t_y \\
         r_w & r_y
      \end{pmatrix}
      =
      \begin{pmatrix}
         t_w & t_y \\
         0 & 1
      \end{pmatrix} ~,
 \end{align}
 which retains $y$ as a comoving coordinate, so the metric becomes
 \begin{align}
   ds^2 & = -(t_w \, dw + t_y \, dy)^2 + \frac{(R')^2}{1 + f} \, (r_w \, dw + r_y \, dy)^2 \nn \\
           &~~~ + R^2 \, d\Omega^2 \\
   & = - t_w^2 \, dw^2 - 2 \, t_w \, t_y \, dw \, dy
      + \left( - t_y^2 + \frac{(R')^2}{1 + f} \right) dy^2 \nn \\
      &~~~ + R^2 \, d\Omega^2 ~.
 \end{align}
 We want $w$ to be a coming null coordinate, i.e. $dw = 0 = d\theta = d\phi$ must give 
$ds = 0$, which leads to
 \begin{align}
     g_{yy}  & = 0 ~~~~\to~~~~ t_y = \frac{-R'}{\sqrt{1 + f}\;}
      \eqlabel{tyRr} \\
   \to~~~~~~ ds^2 & = - t_w^2 \, dw^2
      + 2 \, t_w \, \frac{R'}{\sqrt{1 + f}\;} \, dw \, dy
      + R^2 \, d\Omega^2 ~,
 \end{align}
 where the sign choice is because, on the past null cone (PNC), $t$ must decrease as $r = y$ increases.  Eq (\ref{tyRr}) is a PDE for $t(w, y)$, and its solution will introduce a function of $w$, $\gamma(w)$.

 Applying this transformation to the LT evolution equation \er{LTev}
 \begin{align}
   R_w & = \dot{R} \, t_w + R' \, r_w = \dot{R} \, t_w ~,
 \end{align}
 we find the new evolution equation in the new coordinates, 
 \begin{align}
   R_w & = \pm t_w \sqrt{\frac{2 M}{R} + f + \frac{\Lambda R^2}{3}}\; ~.
   \eqlabel{RwEq}
 \end{align}
 Similarly, by transforming $R_y$ we obtain
 \begin{align}
   R_y & = \dot{R} \, t_y + R' \, r_y
      = - \dot{R} \left( \frac{R'}{\sqrt{1 + f}\;} \right) + R' \\
   \to~~~~~~ R' & = R_y \left( \frac{\sqrt{1 + f}\;}
      {\sqrt{1 + f}\; \mp \sqrt{\frac{2 M}{R} + f + \frac{\Lambda R^2}{3}}\;} \right) ~,
   \eqlabel{RrEq}
 \end{align}
 where (\ref{RwEq}) was used.  The metric now becomes
 \begin{align}
   ds^2 & = - t_w^2 \, dw^2 + \frac{2 \, R_y \, t_w}
      {\left( \sqrt{1 + f}\; \mp \sqrt{\frac{2 M}{R} + f + \frac{\Lambda R^2}{3}}\; \right)} \, dw \, dy \nn \\
        &~~~ + R^2 \, d\Omega^2 ~.
      \eqlabel{LTObsMetric}
 \end{align}
 Up to this point $t(w, y)$ contains an undetermined function of $w$, obtained when integrating (\ref{tyRr}).  We now specify that, at the origin, 
$y = r = 0$, $w$ is the observer's proper time, $w = t$, i.e.
 \begin{align}
   t(w, 0) & = w ~,~~~~~~ w(t, 0) = t ~,~~~~~~ t_w(w, 0) = 1 ~,\nn \\ 
           & w_t(t, 0) = 1 ~.
 \end{align}
 and this will fix the function of integration introduced in solving (\ref{tyRr}).

 Comparing \er{ObsMetric} and \er{LTObsMetric}, it is clear that
 \begin{align}
   A & = t_w ~,\nn \\
   B &= - t_y = \frac{R_y}{\left( \sqrt{1 + f}\; \mp \sqrt{\frac{2 M}{R} + f + \frac{\Lambda R^2}{3}}\; \right)} ~.
 \end{align}
 in agreement with \er{Atw} and \er{CyB}.

 The matter tensor transforms to 
 \begin{align}
   \tilde{T}^{ab} & = T^{cd} \, (J^{-1})_c^a \, (J^{-1})_d^b = 
      \begin{pmatrix}
         \rho / t_w^2 & 0 \\
         0 & 0
      \end{pmatrix} ~,\nn \\ 
   \tilde{T}_{ab} & = T_{cd} \, J^c_a \, J^d_b = 
      \begin{pmatrix}
         \rho \, t_w^2 & \rho \, t_w \, t_y \\
         \rho \, t_w \, t_y & \rho \, t_y^2
      \end{pmatrix} ~.
 \end{align}


\begin{thebibliography}{99}

 \bibitem{KriSac66} J. Kristian \& R.K. Sachs
 (1966) {\it Astrophys. J.} {\bf 143}, 379-99,
 ``Observations in Cosmology".
     %
 \bibitem{ENMSW85} G.F.R. Ellis, S.D. Nel, R. Maartens, W.R. Stoeger, \& A.P. Whitman
 (1985) {\it Phys. Reports} {\bf 124}, 315-417,
 ``Ideal Observational Cosmology". 
     %
 \bibitem{SNME92} W.R. Stoeger, S.D. Nel, R. Maartens \& G.F.R. Ellis 
 (1992), {\it Class. Q. Grav.}, {\bf 9}, 493-507,
 ``The Fluid-Ray Tetrad Formulation of Einstein's Field Equations". 
     %
 \bibitem{SEN92} W.R. Stoeger, G.F.R. Ellis \& S.D. Nel
 (1992a), {\it Class. Q. Grav.}, {\bf 9}, 509-26,
 ``Observational Cosmology: III. Exact Spherically Symmetric Dust Solutions".
     %
 \bibitem{SNE92b} W.R. Stoeger, S.D. Nel \& G.F.R. Ellis
 (1992b), {\it Class. Q. Grav.}, {\bf 9}, 1711-23,
 ``Observational Cosmology: IV. Perturbed Spherically Symmetric Dust Solutions". 

 \bibitem{SNE92c} W.R. Stoeger, S.D. Nel \& G.F.R. Ellis
 (1992c), {\it Class. Q. Grav.}, {\bf 9}, 1725-51,
 ``Observational Cosmology: V. Solutions of the First Order General Perturbation Equations."
     %
 \bibitem{MaaMat94} R. Maartens, D.R. Matravers
 (1994) {\it Class. Q. Grav.} {\bf 11}, 2693-704,
``Isotropic and Semi-Isotropic Observation in Cosmology".
     %
 \bibitem{MHMS96} R. Maartens, N.P. Humphreys, D.R. Matravers, \& W.R. Stoeger
 (1996) {\it Class. Q. Grav.}, {\bf 13}, 253-64.
 ``Inhomogeneous Universes in Observational Coordinates";
 plus Errata in (1996) {\it Class. Q. Grav.}, {\bf 13}, 1689-90.
     %
 \bibitem{AraSto99} M.E. Ara\'{u}jo \& W.R. Stoeger
 (1999) {\it Phys. Rev. D}, {\bf 60}, 104020, 1-7,
 ``Exact Spherically Symmetric Dust Solution of the Field Equations in Observational Coordinates with Cosmological Data Functions";
 plus Errata in (2001) {\it Phys. Rev. D}, {\bf 64}, 049901, 1. 
     %
 \bibitem{AABFS01} M.E. Ara\'{u}jo, R.C. Arcuri, J.L. Bedran, L.R. de Freitas, \& W.R. Stoeger
 (2001) {\it Astrophys. J.} {\bf 549}, 716-20, 
``Integrating Einstein Field Equations in Observational Coordinates with Cosmological Data Functions: Nonflat Friedmann-Lemaitre-Robsertson-Walker Cases".
     %
 \bibitem{ARS01} M.E. Ara\'{u}jo, S.R.M.M. Roveda \& W.R. Stoeger 
(2001) {\it Astrophys. J.} {\bf 560}, 7-14,
 ``Perturbed Spherically Symmetric Dust Solution of the Field Equations in Observational Coordinates with Cosmological Data Functions".
     %
 \bibitem{RibSto03} M.B. Ribeiro \& W.R. Stoeger,
 (2003) {\it Astrophys. J.} {\bf 592}, 1-16,
``Relativistic Cosmology Number Counts and the Luminosity Function".
     %
 \bibitem{AruSto07} M.E. Ara\'{u}jo \& W. R. Stoeger
 (2007) {\tt arXiv:0705.1846 [astro-ph]},
 ``The Angular-Diameter-Distance-Maximum and Its Redshift as Constraints on $\Lambda \neq 0$ FLRW Models"
     %
 \bibitem{AruSto08} M.E. Ara\'{u}jo, W.R. Stoeger, R.C. Arcuri \& M.L. Bedran,
 {\tt arXiv:0807.4193v1 [astro-ph]} (2008),
 ``Solving Einstein Field Equations in Observational Coordinates with Cosmological Data Functions: Spherically Symmetric Universes with Cosmological Constant"
     %
 \bibitem{Tem38} G. Temple
 (1938) {\it Proc. Roy. Soc. London A}, {\bf 168}, 122-48,
 ``New Systems of Normal Coordinates for Relativistic Optics".
     %
 \bibitem{Cel00} M.-N. C\'el\'erier
 (2000) {\it Astron. Astrophys.} {\bf 353}, 63,
 ``Do We Really See a Cosmological Constant in the Supernovae Data?
     %
 \bibitem{TanNam07} M. Tanimoto \& T. Nambu,
 (2007) {\it Class. Q. Grav.},  {\bf 24}, 3843,
 ``Luminosity Distance-Redshift Relation for the LTB Solution Near the Centre".
     %
 \bibitem{MHE97} N. Mustapha, C. Hellaby \& G.F.R. Ellis
 (1997) {\it Mon. Not. Roy. Astron. Soc.} {\bf 292}, 817-30,
 ``Large Scale Inhomogeneity Versus Source Evolution: Can We Distinguish Them Observationally?". 
     %
 \bibitem{MBHE98} N. Mustapha, B.A.C.C. Bassett, C. Hellaby \& G.F.R. Ellis
 (1998) {\it Class. Q. Grav.} {\bf 15}, 2363-79,
 ``The Distortion of the Area Distance-Redshift Relation in Inhomogeneous Isotropic Universes". 
     %
 \bibitem{Hel01} C. Hellaby,
 (2001) {\it Astron. Astrophys.} {\bf 372}, 357-363,
 ``Multicolour Observations, Inhomogeneity and Evolution".
     %
 \bibitem{Hel06} C. Hellaby
 (2006) {\it Mon. Not. Roy. Astron. Soc.} {\bf 370}, 239-244,
 ``The Mass of the Cosmos".
     %
 \bibitem{Lu06} T. H.-C. Lu
 (2006) MSc Thesis, University of Cape Town.
     %
 \bibitem{LuHel07} T. H.-C. Lu \& C. Hellaby,
 (2007) {\it Class. Q. Grav.}, {\bf 24}, 4107, 
 ``Obtaining the Spacetime Metric from Cosmological Observations".
     %
 \bibitem{McCHel08} M. L. McClure \& C. Hellaby
 (2008) {\it Phys. Rev. D} {\bf 78}, 044005, 1-17,
 ``Determining the Metric of the Cosmos: Stability, Accuracy, and Consistency".
     %
 \bibitem{Ish04}  M. Ishak (2004)
 {\it Phys. Rev. D} {\bf 69}, 124027, 
 ``On Perfect Fluid Models in Non-Comoving Observational Spherical Coordinates".
     %
 \bibitem{BisHai96} N. Bishop, and P. Haines (1996)
 {\it Quaestiones Mathematicae} {\bf 19}, 259,
 ``Observational Cosmology and Numerical Relativity".
     %
 \bibitem{Eth33} I.M.H. Etherington
 (1933) {\it Phil. Mag. VII} {\bf 15}, 761-73,
 ``On the Definition of Distance in General Relativity";
 reprinted in (2007) {\it Gen. Rel. Grav.} {\bf 39}, 1055.
     %
 \bibitem{Penr66} R. Penrose
 (1966) in {\it Perspectives in Geometry and Relativity: Essays in Honour of Vaclav Hlavaty},
 Ed B. Hoffman, Indiana University Press, pp 259-74, 
 ``General Relativistic Energy Flux and Elementary Optics".
     %
 \bibitem{Ell71} G.F.R. Ellis
 (1971) in {\it General Relativity and Cosmology},
 Proc. Int. School of Physics ``Enrico Fermi" (Varenna), Course XLVII, 
 Ed. R. K. Sachs (Academic Press), pp 104-79,
 ``Relativistic Cosmology".
     %
 \bibitem{Lem33} G. Lema\^{\i}tre
 (1933) {\it Ann. Soc. Sci. Bruxelles} {\bf A53}, 51-85,
 ``L'Univers en Expansion";
 reprinted in (1997) {\it Gen. Rel. Grav.} {\bf 29}, 641-80.
     %
 \bibitem{Tol34} R.C. Tolman
 (1934) {\it Proc. Nat. Acad. Sci.} {\bf 20}, 169-76,
 ``Effect of Inhomogeneity on Cosmological Models";
 reprinted in (1997) {\it Gen. Rel. Grav.} {\bf 29}, 935-43.
     %
 \end{thebibliography}
 \end{document}